\documentclass[draftcls,onecolumn]{IEEEtran}
\usepackage{amsmath,amssymb,cite,color}

\newtheorem{theorem}{Theorem}
\newtheorem{lemma}{Lemma}

\newtheorem{remark}{Remark}

\title{Analysis of Nakamoto Consensus, Revisited}
\author{
\IEEEauthorblockN{Jianyu Niu\IEEEauthorrefmark{1}, Chen Feng\IEEEauthorrefmark{1}, Hoang Dau\IEEEauthorrefmark{2}, Yu-Chih Huang\IEEEauthorrefmark{3}, Jingge Zhu\IEEEauthorrefmark{4}}

\IEEEauthorblockA{\IEEEauthorrefmark{1}The University of British Columbia (Okanagan Campus)}

\IEEEauthorblockA{\IEEEauthorrefmark{2}RMIT University}
\IEEEauthorblockA{\IEEEauthorrefmark{3}National Taipei University}

\IEEEauthorblockA{\IEEEauthorrefmark{4}The University of Melbourne}
}

\begin{document}

\maketitle

\section{Introduction}

In the Bitcoin white paper~\cite{Nakamoto}, Nakamoto proposed a very simple Byzantine fault tolerant consensus algorithm that is also known as Nakamoto consensus. Despite its simplicity, some existing analysis of Nakamoto consensus appears to be long and involved. In this technical report, we aim to make such analysis simple and transparent so that we can teach senior undergraduate students and graduate students in our institutions. This report is largely based on a 3-hour tutorial given by one of the authors in June 2019~\cite{Talk}.

\section{System Model}

We closely follow the notations in~\cite{Tse}. Let $\mathcal{N}$ denote the set of participating nodes in the network.
Each node $n \in \mathcal{N}$ has $p_n$ fraction of total hashing power so that it mines new blocks at a rate
of $p_n f$ blocks per second, where $f$ is the total mining rate.\footnote{We assume constant mining difficulty here.} There are two types of nodes: honest nodes
who strictly follow the protocol and adversarial nodes who may deviate from the protocol. The set of honest
nodes (resp., adversarial nodes) is denoted by $\mathcal{H}$ (resp., $\mathcal{Z}$). 
The adversarial nodes control $\beta$ fraction of total hashing power, i.e., $\sum_{n \in \mathcal{Z}} p_n 
= \beta$. The honest nodes control $1 - \beta$ fraction of total hashing power, i.e., $\sum_{n \in \mathcal{H}} p_n 
= 1 - \beta$. If a block is mined by an honest node (resp., adversarial node), we call it an honest block 
(resp., adversarial block).
A block's height is its parent block's height plus one. (The height of the genesis block is set to $0$.)

Mining at each node $n \in \mathcal{N}$ is modeled by a Poisson process with rate $p_n f$ as done in the Bitcoin
white paper. Hence, the aggregated mining process of the honest nodes (resp., adversarial nodes) is a Poisson process with rate $(1 - \beta) f$ (resp., $\beta f$). Without loss of generality, we can assume a \emph{single} adversarial node with 
$\beta$ fraction of hashing power, and we call this node the adversary.

We assume a bounded network delay of $\Delta$ seconds for honest nodes. That is, whenever an honest node mines
a new block, it takes up to $\Delta$ seconds for the block to reach all other honest nodes. We assume a zero delay
from honest nodes to the adversary. That is, whenever an honest node mines a new block, the adversary receives it immediately. These assumptions make the adversary even more powerful in terms of network
communication.


Next, we discretize the above continuous-time model into a discrete-time model in a way that generalizes 
the discretization procedure in~\cite{Tse}. We divide $\Delta$ seconds into $\tau$ rounds so that each round
in our model corresponds to $\frac{\Delta}{\tau}$ seconds. More specifically, round $0$ corresponds to
the time interval $[0, \frac{\Delta}{\tau})$, and round $r$ corresponds to the interval 
$[r\frac{\Delta}{\tau}, (r+1) \frac{\Delta}{\tau})$.
When $\tau = 1$, our model reduces to the discrete-time model
in~\cite{Tse}. On the other hand, when $\tau \to \infty$, our model approaches the continuous-time model in~\cite{UIUC}.
In this sense, our model provides a unified treatment. 

Following~\cite{Tse, Pass, Garay}, we assume that blocks can only be mined at the beginning of each round.
That is, if new blocks are mined in round $r$ with the interval $[r\frac{\Delta}{\tau}, (r+1) \frac{\Delta}{\tau})$, 
we will set their generation time to be the beginning of round $r$ (i.e., $r\frac{\Delta}{\tau}$).
Note that such an approximation tends to be accurate as $\tau \to \infty$.
Let $H[r]$ and $Z[r]$ be the number of blocks mined by the honest nodes and by the adversary, respectively, 
in round $r$. Clearly, $H[r]$ and $Z[r]$ are independent Poisson random variables with means
$(1 - \beta) f \frac{\Delta}{\tau}$ and $\beta f \frac{\Delta}{\tau}$, respectively.
In addition, the sequences $\{ H[0], H[1], \ldots \}$ and $\{ Z[0], Z[1], \ldots \}$ are independent
of each other and independent across rounds.
Note that the $H[r]$ honest blocks mined at the beginning of round $r$ will reach all the honest nodes in the network
by the end of round $r + \tau - 1$, since it takes $\tau$ rounds to broadcast any honest block.
On the other hand, the $Z[r]$ adversarial blocks can be kept in private until the adversary decides
to transmit any of them in later rounds. Once transmitted, any adversarial block will reach all
the honest nodes within $\tau$ rounds.\footnote{In some implementation of Bitcoin, certain blocks of small heights may be discarded by honest nodes so that these blocks won't be broadcasted to the entire network. In our system model, we assume that any block will be broadcasted unless it is kept in private.}

Under the above system model, Nakamoto consensus can be described as follows.
\begin{itemize}
    \item At each round $r$, an honest node attempts to mine new blocks on top of the longest chain it observes by the end of round $r-1$ (where ties can be broken arbitrarily). This is often referred to as the longest chain rule.
    \item At each round $r$, an honest node confirms a block if the longest chain it adopts contains the block as
    well as at least $k$ other blocks of larger heights. This is sometimes referred to as the $k$-deep confirmation rule.
\end{itemize}

\begin{table}[tp]
\centering
\caption{Key notations in the system model}
\label{table:1}
\begin{tabular}{ |c|l| } 
 \hline
 $\Delta$ & Upper bound on the network communication delay (in seconds) \\ 
 $f$ & Total mining rate (in blocks per second)  \\ 
 $\beta$ & Fraction of adversarial mining rate \\ 
 $H[r]$ & Number of honest blocks mined in round $r$ \\ 
 $Z[r]$ & Number of adversarial blocks mined in round $r$ \\ 
 \hline
\end{tabular}
\end{table}

Next, let us make an observation that will be used in our analysis later.

\begin{lemma}\label{lem:height}
If an honest block of height $\ell$ is mined at the beginning of round $r$, then every honest node observes a chain of length at least $\ell$ by the end of round $r + \tau - 1$.
\end{lemma}

\begin{IEEEproof}
First, this honest block will reach all the honest nodes 
by the end of round $r + \tau - 1$ as we discussed before. 
Second, its parent block
(no matter honest or adversarial) will reach all the honest nodes 
by the end of round $r + \tau - 1$. This argument applies to 
all of its ancestor blocks. Hence, by the end of round $r + \tau - 1$, every honest node will observe
a chain consisting of this block, its ancestor blocks, as well as new (honest or adversarial) blocks mined on top of this block. If there are no such new blocks, the chain length is $\ell$. Otherwise, the chain length is greater than $\ell$.
\end{IEEEproof}



\section{Effective Rounds and Liveness}
A round $r$ is called an effective round (ER) if there is some honest block mined in round $r$ and there is 
no honest block mined in the previous $\tau - 1$ rounds. 
By effective, we mean two things: 1) at least one honest block is successfully mined
in round $r$ and 2) the longest chain (among all the honest nodes) will be increased.
When $\tau = 1$, a round $r$ is an ER if and only if $H[r] \ge 1$. When $\tau > 1$, a round $r$ is an ER if and only if $H[r] \ge 1$ and $H[r'] = 0$ for all $r' \in \{ r - (\tau - 1), \ldots, r -1 \}$.
For convenience, we assume that $H[r'] = 0$ for all $r' < 0$. 


\begin{lemma}\label{lem:ER}
Honest blocks mined in distinct ERs have different heights.
\end{lemma}

\begin{IEEEproof}
Suppose for contradiction that two honest blocks $B$ and $B'$ of height $\ell$ are mined in round $r$ and $r'$ respectively. Without loss of generality, assume that $r < r'$. We have $r' \ge
r + \tau$, because otherwise $r'$ cannot be an ER. By Lemma~\ref{lem:height}, every honest node observes
a chain of length at least $\ell$ by the end of round $r'-1$ (or even earlier). 
Therefore, no honest node will mine a new block $B'$ of height $\ell$ in round $r'$.
\end{IEEEproof} 


Next, we introduce an indicator random variable $X[r]$ for whether round $r$ is an ER, i.e., $X[r] = 1$ when
round $r$ is an ER and $X[r] = 0$ otherwise. Note that $\Pr( X[r] = 1) \ge e^{-(1 - \beta) f \frac{\Delta}{\tau} (\tau - 1)} \left( 1 - e^{-(1 - \beta) f \frac{\Delta}{\tau}} \right)$, where the equality holds when $r \ge \tau$.
For convenience, we write $X[r, r'] \triangleq X[r] + X[r+1] + \cdots + X[r']$. This notation applies to 
other random variables as well, such as $\{H[r]\}$ and $\{Z[r]\}$.

\begin{lemma}\label{lem:expected_ER}
Let $\gamma = e^{-(1 - \beta) f \frac{\Delta}{\tau} (\tau - 1)} \left( 1 - e^{-(1 - \beta) f \frac{\Delta}{\tau}} \right)$. In a time interval of $s$ consecutive rounds, the expected number of ERs is at least $\gamma s$.
\end{lemma}

\begin{IEEEproof}
The number of ERs in a time interval of $s$ consecutive rounds starting from round $r$ is given by
$X[r, r + s - 1]$.
Hence, we have
\begin{equation}
 E \left( X[r, r + s - 1] \right) = E \left( X[r]  \right) + \cdots + E \left(  X[r + s - 1] \right) \ge \gamma s,
\end{equation}
where the equality holds when $r \ge \tau$.
\end{IEEEproof} 

\begin{lemma}\label{lem:bound_ER}
For any positive integer $m$, in a time interval of $\tau m$ consecutive rounds starting from round $r$,  the number of ERs has the following Chernoff-type bound: For $0 < \delta < 1$,
\begin{equation}
   \Pr( X[r, r + \tau m - 1]  \le (1 - \delta) \gamma \tau m) \le  e^{-\Omega\left(\delta^2 \gamma  m \right)}. 
\end{equation}
\end{lemma}

\begin{IEEEproof}
Let $X^{(j)} = \sum_{i = 0}^{m - 1} X[r + j + i\tau]$. Then, $X[r, r + \tau m - 1] = X^{(0)} + \cdots + X^{(\tau - 1)}$. Our key observation is
that $\{ X[r+j], X[r + j + \tau], \ldots, X[r + j + (m-1)\tau] \}$ are independent
random variables, because $X[r]$ is a function of
$\{H[r - (\tau -1)],\ldots, H[r]\}$. By (a slightly modified version of) Lemma~\ref{lem:expected_ER},
we have $E\left( X^{(j)} \right) \ge \gamma m$.
By Lemma~\ref{lem:key_step}, we have
$\Pr\left( X[r, r + \tau m - 1] \le (1 - \delta)\gamma \tau m \right) \le  e^{-\Omega\left(\delta^2 \gamma  m \right)}$.
\end{IEEEproof} 

\begin{theorem}[Chain growth]
If an honest node observes a chain of length $\ell$ at the beginning of round $r$,
then at the beginning of round $r + \tau(m + 2) - 1$, every honest node observes a chain
of length at least $\ell + (1 - \delta) \gamma \tau m$, except for $e^{-\Omega\left({\delta^2 \gamma m}\right)}$ probability.
\end{theorem}

\begin{IEEEproof}
First, by Lemma~\ref{lem:height}, every honest node observes a chain
of length at least $\ell$ at the beginning of round $r + \tau$. 
Next, consider a time interval of $\tau m$ consecutive rounds
starting from round $r + \tau$. By Lemma~\ref{lem:bound_ER}, we have at least $(1 - \delta) \gamma \tau m$ ERs
from round $r + \tau$ to round $r + \tau + \tau m - 1$,
except for $e^{-\Omega\left({\delta^2 \gamma m}\right)}$ probability.
We take one honest block from each ER. By Lemma~\ref{lem:ER}, these honest blocks
have different heights, all greater than $\ell$. 
In particular, the largest height of these blocks
is at least $\ell + (1 - \delta) \gamma \tau m$. By Lemma~\ref{lem:height},
at the beginning of round $r + 2 \tau + \tau m - 1$, every honest node observes 
a chain of length at least $\ell + (1 - \delta) \gamma \tau m$.
\end{IEEEproof} 

\begin{theorem}[Chain quality]
Suppose $\gamma > (1 + \delta) \beta f \frac{\Delta}{\tau}$. In a time interval of $\tau m$ consecutive rounds starting from round $0$, in the longest chain among
honest nodes, the fraction of honest blocks is at least $1 - (1+\delta) \frac{\beta f \frac{\Delta}{\tau}}{\gamma}$ except for $e^{-\Omega\left( \delta^2
\min\{\beta f \Delta, \gamma \}m \right)}$ probability.
\end{theorem}

\begin{IEEEproof}
We let $s = \tau m$ for convenience. On the one hand, $Z[0, s-1]$ is the number of adversarial blocks from round $0$ to round $s-1$, which is a Poisson 
random variable with mean $\beta f \frac{\Delta}{\tau} s$. By Lemma~\ref{lem:Poisson}, $Z[0, s-1] < 
(1 + \delta_Z) \beta f \frac{\Delta}{\tau} s$ except for $e^{-\Omega\left( \delta^2
\beta f \Delta m \right)}$ probability.
On the other hand, $X[0, s-1]$ is the number of ERs from round $0$ to round $s-1$.
By Lemma~\ref{lem:bound_ER}, $X[0, s-1] > (1 - \delta_X) \gamma s$
except for $e^{-\Omega\left({\delta_X^2 \gamma m}\right)}$ probability.
Hence, by the end of round $s-1$, some honest node observes
an honest block of height at least $(1 - \delta_X) \gamma s$. In other words,
by the end of round $s-1$,
the length of the longest chain among honest nodes, denoted by $L(s)$, is at least $(1 - \delta_X) \gamma s$.
The honest fraction is smallest if all the adversarial blocks belong to the longest chain of length $L(s)$. 
That is, the honest fraction is at least $\frac{L(s) - Z[0, s-1]}{L(s)}$, which is lower bounded by $\frac{X[0, s-1] - Z[0, s-1]}{X[0, s-1]}$.
Finally, by setting $\delta_Z = \delta_X = \delta/4$ and noticing $\frac{1 + \delta/4}{1 - \delta/4} < 1 + \delta$, we have
\begin{equation}
\frac{X[0, s-1] - Z[0, s-1]}{X[0, s-1]} > 1 - \frac{1 + \delta_Z}{1 - \delta_X} \frac{\beta f \frac{\Delta}{\tau} s}{\gamma s} > 1 - (1+\delta) \frac{\beta f \frac{\Delta}{\tau}}{\gamma}, 
\end{equation}
except for $e^{-\Omega\left( \delta^2
\min\{\beta f \Delta, \gamma \}m \right)}$ probability.
\end{IEEEproof}

Finally, we would like to point out that chain growth and chain quality---when putting together---imply \emph{livenss}, which states that every valid transaction will be eventually confirmed by honest nodes with high probability.

\section{Uniquely Effective Rounds and Safety}

A round $r$ is called a uniquely effective  round (UER) if there is exactly one honest block mined in round $r$, and there is 
no honest block mined in the previous and next $\tau - 1$ rounds. 
By uniquely effective, we mean two things: 1) a unique honest block is successfully mined
in round $r$ and 2) the honest block has a unique height among all other honest blocks, as stated in Lemma~\ref{lem:UER}.
When $\tau = 1$, a round $r$ is a UER if and only if $H[r] = 1$. When $\tau > 1$, a round $r$ is a UER if and only if $H[r] = 1$ and $H[r'] = 0$ for all $r' \in \{ r - (\tau - 1), \ldots, r -1, r+1, 
\ldots, r + (\tau - 1)\}$. 

\begin{lemma}\label{lem:UER}
Suppose that an honest block $B$ of height $\ell$ is mined in a UER. Then
$B$ is the only honest block of height $\ell$.
\end{lemma}

\begin{IEEEproof}
Suppose for contradiction that two honest blocks $B$ and $B'$ of height $\ell$ are mined in round $r$ and $r'$ respectively. Since 
round $r$ is a UER, we have $r' \ge r + \tau$ or $r' \le r - \tau$. If $r' \ge r + \tau$, 
by Lemma~\ref{lem:height}, every honest node observes a chain of length at least $\ell$ 
by the end of round $r'-1$ (or even earlier). Therefore, no honest node
will mine a new block of height $\ell$ in round $r'$, leading to a contradiction. Similarly, if $r' \le r - \tau$, every honest node observes a chain of length at least $\ell$ 
by the end of round $r-1$ (or even earlier), leading to a contradiction.
\end{IEEEproof}

Next, we introduce an indicator random variable $Y[r]$ for whether round $r$ is a UER, i.e., $Y[r] = 1$ when
round $r$ is a UER and $Y[r] = 0$ otherwise. Note that $\Pr( Y[r] = 1) \ge (1 - \beta) f \frac{\Delta}{\tau} e^{-(1 - \beta) f \frac{\Delta}{\tau} (2\tau - 1)}$, where the equality holds when $r \ge \tau$.

\begin{lemma}\label{lem:expected_UER}
Let $\eta = (1 - \beta) f \frac{\Delta}{\tau} e^{-(1 - \beta) f \frac{\Delta}{\tau} (2\tau - 1)}$. In a time interval of $s$ consecutive rounds, the expected number of UERs is at least $\eta s$.
\end{lemma}

\begin{IEEEproof}
The number of UERs in a time interval of $s$ consecutive rounds starting from round $r$ is given by
$Y[r, r + s - 1]$.
Hence, we have
\begin{equation}
   E \left( Y[r, r + s - 1] \right) = E \left( Y[r]  \right) + \cdots + E \left(  Y[r + s - 1] \right)\ge \eta s, 
\end{equation}
where the equality holds when $r \ge \tau$.
\end{IEEEproof}

\begin{lemma}\label{lem:bound_UER}
For any positive integer $m$, in a time interval of $(2 \tau - 1)m$ consecutive rounds starting from round $r$,  the number of UERs has the following Chernoff-type bound: For $0 < \delta < 1$,
\begin{equation}
  \Pr( Y[r, r + (2 \tau - 1)m - 1]  \le (1-\delta) \eta (2 \tau - 1)m) \le e^{-\Omega\left({\delta^2 \eta m}\right)}.  
\end{equation}
\end{lemma}

\begin{IEEEproof}
Let $Y^{(j)} = \sum_{i = 0}^{m - 1} Y[r + j + i(2\tau-1)]$. Then, $Y[r, r + (2 \tau - 1)m - 1] = Y^{(0)} + \cdots + Y^{(2\tau - 2)}$. Our key observation is that $\{ Y[r+j], Y[r + j + (2\tau-1)], \ldots, Y[r + j + (m-1)(2\tau-1)] \}$ are independent
random variables. By (a slightly modified version of) Lemma~\ref{lem:expected_UER},
we have $E\left( Y^{(j)} \right) \ge \eta m$.
By Lemma~\ref{lem:key_step}, we have $\Pr\left( Y[r, r + (2 \tau - 1)m - 1] \le (1 - \delta)\eta (2\tau -1) m \right) \le  e^{-\Omega\left(\delta^2 \eta  m \right)}$.
\end{IEEEproof}

\begin{lemma}\label{lem:bound2_UER}
Suppose $\eta > (1 + \delta) \beta f \frac{\Delta}{\tau}$. 
In a time interval of $(2 \tau - 1) m$ consecutive rounds starting from round $r$, 
the number of UERs is greater than the number of adversarial blocks except
for $e^{-\Omega\left(\delta^2 \min\{ \eta, \beta f \Delta  \}  m \right)}$ probability. That is, 
\begin{equation}
 \Pr\left( Y[r, r + (2 \tau - 1) m - 1] \le Z[r, r + (2 \tau - 1) m - 1] \right) \le e^{-\Omega\left(\delta^2 \min\{ \eta, \beta f \Delta  \}  m \right)}.   
\end{equation}
\end{lemma}

\begin{IEEEproof}
We let $s = (2 \tau - 1) m$ for convenience. Let $Y = Y[r]  + \cdots + Y[r + s - 1]$ and $Z = Z[r] + \cdots + Z[r + s - 1]$.
Then, by Lemma~\ref{lem:bound_UER}, $Y > (1 - \delta_Y)\eta s$ except for $e^{-\Omega\left(\delta_Y^2 \eta  m \right)}$ probability.
Similarly, by Lemma~\ref{lem:Poisson}, $Z < (1 + \delta_Z) \beta f \frac{\Delta}{\tau} s$ except for 
$e^{-\Omega\left(\delta_Z^2 \beta f \Delta  m \right)}$ probability.
By setting $\delta_Y = \delta_Z = \delta/4$ and noticing $\frac{1 + \delta/4}{1 - \delta/4} < 1 + \delta$, we have
$(1 - \delta_Y)\eta > (1 + \delta_Z) \beta f \frac{\Delta}{\tau}$. Therefore, $Y > Z$
except for $e^{-\Omega\left(\delta^2 \min\{ \eta, \beta f \Delta  \}  m \right)}$ probability.
\end{IEEEproof}

\begin{remark}
Note that the condition $\eta > (1 + \delta) \beta f \frac{\Delta}{\tau}$ is equivalent to $f \frac{\Delta}{\tau} (2\tau - 1) < \frac{1}{1 - \beta} \ln\left( \frac{1 - \beta}{\beta} \frac{1}{1 + \delta} \right)$.
This implies $\beta < 0.5$. When $\tau = 1$, the condition says $f \Delta < \frac{1}{1 - \beta} \ln\left( \frac{1 - \beta}{\beta} \frac{1}{1 + \delta} \right)$.
When $\tau \to \infty$, the condition says $f \Delta < \frac{1}{2} \frac{1}{1 - \beta} \ln\left( \frac{1 - \beta}{\beta} \frac{1}{1 + \delta} \right)$.
\end{remark}

\begin{theorem}[Safety]\label{thm:safety}
Suppose $\eta > (1 + \delta) \beta f \frac{\Delta}{\tau}$. If $B$
and $B'$ are two distinct blocks of the same height, then they cannot be
both confirmed, each by an honest node. This property holds, regardless of adversarial action, except for $e^{-\Omega\left(\delta^2 \min\{ \frac{\eta}{f \Delta}, \beta   \} k \right)}$ probability.
\end{theorem}

\begin{IEEEproof}
Consider the event $\mathcal{E}$ that ``$B$ and $B'$ of the same height are both confirmed, each by an honest node." We will show that
this event happens with probability at most $e^{-\Omega\left(\delta^2 \min\{ \frac{\eta}{f \Delta}, \beta   \} k \right)}$, regardless
of adversarial action.
Let $r$ (resp., $r'$) be the smallest round at the beginning of which $B$ (resp., $B'$) is confirmed. Without loss of generality, we assume that $r \ge r'$. Let $B_1$ be the most recent ancestor of $B$ and $B'$. That is, there are two disjoint subchains mined on top of $B_1$, one containing $B$ and the other containing $B'$. Let $B_0$ be the most recent \emph{honest} ancestor of
$B$ and $B'$. Note that $B_0$ can be $B_1$ (if $B_1$ is honest) or the genesis block. Suppose that $B_0$ is mined (by some honest node)
at the beginning of round $r_0$. For convenience, we assume that the genesis block is mined at the beginning of round $0$. This makes $r_0$ well defined. We next define the following two events:
\begin{itemize}
    \item $\mathcal{E}_1(r_0, r)$: At the beginning of round $r$, there are two disjoint subchains mined on top of $B_1$, each containing at least $k + 1$ blocks mined from round $r_0$ to round $r$; 
    \item $\mathcal{E}_2(r_0, r)$: $Y[r_0 + \tau, r - \tau] \le Z[r_0, r]$.
\end{itemize}
We will show that $\mathcal{E} \subseteq \mathcal{E}_1(r_0, r)  \subseteq \mathcal{E}_2(r_0, r)$, regardless of 
adversarial action. 

\begin{itemize}
    \item $\mathcal{E} \subseteq \mathcal{E}_1(r_0, r)$: At the beginning of round $r$, one subchain contains $B$ as well as $k$ blocks mined on top of $B$ (due to the $k$-deep confirmation rule). Similarly, the other subchain contains $B'$ as well as $k$ other blocks on top of $B'$. These blocks cannot be mined before $r_0$, because $B_0$ is an honest block. 
    \item $\mathcal{E}_1(r_0, r) \subseteq \mathcal{E}_2(r_0, r)$: We will show that whenever there is a unique honest block of height $\ell$ mined in a UER between round $r_0 + \tau$ and round $r - \tau$, there must be a ``matching"  adversarial block of height $\ell$ mined between round $r_0$ and round $r$. To see this, 
    suppose that an honest block $B^*$ is mined in a UER without a matching
    adversarial block. By Lemma~\ref{lem:height}, $B^*$ has a larger height than $B_0$. 
    On the one hand, if $B^*$ has a smaller height than $B$, 
    then $B^*$ must be an honest ancestor of $B$ and $B'$, because $B^*$ is the only block at its height. This contradicts with the fact that $B_0$ is the most recent honest ancestor. 
    On the other hand, if $B^*$ has a larger height than $B$, then both subchains will contain $B^*$ at the beginning of round $r$.
    This is because $B^*$---the only block at its height---will reach all the honest nodes by the end of round $r-1$. As a result, 
    the subchain with $B$ will certainly contain $B^*$ and so the height of $B^*$ is at most the height of $B$ plus $k$. Similarly, the subchain with $B'$ will contain $B^*$, since there are at least $k$ blocks on top of $B'$. This leads to a contradiction.
\end{itemize}
By (a slightly modified version of) Lemma~\ref{lem:bound2_UER}, for any 
    given $r_0$ and $r$, we have  
    \begin{equation}
       \Pr(\mathcal{E}_2(r_0, r)) \le e^{-\Omega\left(\delta^2 \min\{ \eta, \beta f \Delta  \}  \frac{r - r_0 + 1}{2 \tau - 1}  \right)}.  
    \end{equation}
    
    Finally, we will bound $r - r_0$ and complete the proof. We claim that
    \begin{equation}
        r - r_0 + 1 > \frac{2k + 2}{(1+\delta) f \frac{\Delta}{\tau}}
    \end{equation}
    except for $e^{-\Omega\left(\delta^2 k \right)}$ probability, regardless of adversarial action.
    To see this, recall that $\mathcal{E}_1(r_0, r)$ states that two subchains contain at least $2k + 2$ blocks. Hence, $r - r_0 + 1$ is smallest if all the mined blocks from round $r_0$ to round $r$ (the number 
    of which is $H[r_0, r] + Z[r_0, r]$) belong to
    these two subchains. By Lemma~\ref{lem:Poisson}, 
    \begin{equation}
        \Pr\left( H[r_0, r] + Z[r_0, r] \ge (1+\delta) f \frac{\Delta}{\tau} (r - r_0 + 1)  \right) \le e^{-\delta^2 f \frac{\Delta}{\tau} (r - r_0 + 1)/3}.
    \end{equation}
    So, if we set $r - r_0 + 1 = \frac{2k + 2}{(1+\delta) f \frac{\Delta}{\tau}}$, then we have 
    \begin{equation}
        \Pr\left( H[r_0, r] + Z[r_0, r] \ge 2k + 2  \right)  \le e^{-\delta^2 \frac{(2k + 2)}{1 + \delta}/3}.
    \end{equation}
    This proves our claim.
    
    Define the event $\mathcal{D}$ as $r - r_0 + 1 > \frac{2k + 2}{(1+\delta) f \frac{\Delta}{\tau}}$. 
    Then, $\Pr(\mathcal{D}^c) \le e^{-\Omega\left(\delta^2 k \right)}$, where $\mathcal{D}^c$ is the complement of $\mathcal{D}$. Therefore, for any adversarial action, we have
    \begin{align}
        \Pr(\mathcal{E}) &= \Pr(\mathcal{D}^c) \Pr(\mathcal{E} | \mathcal{D}^c) + \Pr(\mathcal{D}) \Pr(\mathcal{E} | \mathcal{D}) \\
        &\le \Pr(\mathcal{D}^c) + \Pr(\mathcal{D}) \Pr(\mathcal{E}_2(r_0, r) | \mathcal{D}) \\
        &\le \Pr(\mathcal{D}^c) + \Pr(\mathcal{E}_2(r_0, r) | \mathcal{D}) \\
        &\le e^{-\Omega\left(\delta^2 \min\{ \frac{\eta}{f \Delta}, \beta   \} k \right)}
    \end{align}
    where the last inequality follows from $k \ge \min\{ \frac{\eta}{f \Delta}, \beta   \} k$.
\end{IEEEproof}

Finally, we would like to point out that our safety property stated in Theorem~\ref{thm:safety} is equivalent to the common-prefix property in 
the previous analysis, such as \cite{Garay, Tse}.

\section{Discussion}

The analysis of Nakamoto consensus was started by Garay, Kiayias and Leonardos in their landmark work \cite{Garay},
which was later refined by Bagaria et. al. \cite{Tse} in the context of parallel chains. Both papers only considered the case of $\tau = 1$. 
The extension to the case of $\tau > 1$ was presented by Pass, Seeman and shelat\footnote{Dr. abhi shelat often writes his name in lower-case and so we follow his style.} \cite{Pass}, which was later
refined by Kiffer, Rajaraman and shelat \cite{Kiffer} as well as Zhao \cite{Zhao} via Markov chain analysis. Our analysis is simpler and more transparent than the previous analysis in that it introduces two events $\mathcal{E}_1(r_0, r)$ and $\mathcal{E}_2(r_0, r)$ explicitly and avoids the use of Markov chains.

At the final stage of completing this report, we notice an independent work by Ling Ren \cite{UIUC}, which focuses on a continuous-time model instead of a discrete-time model. His elegant analysis can be viewed as a counterpart of our analysis. For instance, his safety condition $g^2 \alpha > (1 + \delta) \beta$ is as tight as our safety condition $\eta > (1 + \delta) \beta f \frac{\Delta}{\tau}$ as $\tau \to \infty$.\footnote{We also notice that the safety condition $g \alpha > (1 + \delta) (\alpha + \beta)/2$ in his first version is not as tight as ours.} We will leave it for future work to carefully compare his analysis with ours.

\appendix

\begin{lemma}[Chernoff bounds]\label{lem:Chernoff}
Let $X = \sum_{i = 1}^n X_i$ be the sum of $n$ independent indicator random variables with $E(X_i) = p_i$. Let $\mu = E(X) = \sum_{i = 1}^n p_i$. Then, for $0 < \delta < 1$, $\Pr\left( X \ge (1 + \delta) \mu \right) \le e^{-\delta^2 \mu / 3}$ and $\Pr\left( X \le (1 - \delta) \mu \right) \le e^{-\delta^2 \mu / 2}$.
\end{lemma}

The proof of Lemma~\ref{lem:Chernoff} is elementary. See, e.g., the proof for Theorem~4.5 in~\cite{Book}.

\begin{lemma}[Chernoff bound for a sum of dependent random variables]\label{lem:key_step}
Let $T$ be a positive integer. Let $X^{(j)} = \sum_{i = 0}^{n-1} X_{j + iT}$ be the sum of $n$ independent indicator random variables and $\mu_j = E\left( X^{(j)} \right)$ for $j \in \{1, \ldots, T\}$. Let $X = X^{(1)} + \cdots + X^{(T)}$. 
Let $\mu = \min_j \{ \mu_j \}$. Then, for $0 < \delta < 1$, $\Pr\left( X \le (1 - \delta) \mu T \right) \le e^{-\delta^2 \mu / 2}$.
\end{lemma}

\begin{IEEEproof}
Let $\bar{X} = \frac{X}{T} = \frac{1}{T} \sum_{j = 1}^T X^{(j)}$. Then, for any $t < 0$, we have
\begin{equation}
   \Pr\left( X \le (1 - \delta) \mu T \right) = \Pr\left( \bar{X} \le (1 - \delta) \mu \right) \le \frac{E(e^{t \bar{X}})}{e^{t (1 - \delta) \mu}}. 
\end{equation}
Note that $\exp(\cdot)$ is a convex function, we use Jensen's inequality to obtain $E(e^{t \bar{X}}) \le \frac{1}{T} \sum_{j = 1}^T E\left(e^{t X^{(j)}}\right)$.
Hence,
\begin{equation}
  \Pr\left( X \le (1 - \delta) \mu T \right) \le \frac{1}{T} \sum_{j = 1}^T 
\frac{E\left(e^{t X^{(j)}}\right)}{e^{t (1 - \delta) \mu}} \le 
\frac{1}{T} \sum_{j = 1}^T 
\frac{E\left(e^{t X^{(j)}}\right)}{e^{t (1 - \delta) \mu_j}},  
\end{equation}
where the last inequality comes from the fact that $\mu_j \ge \mu$ for all $j$.
Setting $t = \ln(1 - \delta) < 0$ and following the footsteps in the proof of Theorem~4.5 in~\cite{Book}, we have
$\frac{E\left(e^{t X^{(j)}}\right)}{e^{t (1 - \delta) \mu_j}} \le e^{- \delta^2 \mu_j/2}$
for all $j$. Finally, we note that $e^{- \delta^2 \mu_j/2} \le e^{- \delta^2 \mu/2}$ for all $j$ and this completes the proof.
\end{IEEEproof}

\begin{lemma}[Chernoff bounds for Poisson random variables]\label{lem:Poisson}
Let $X$ be a Poisson random variable with mean $\mu$.  Then, for $0 < \delta < 1$, $\Pr\left( X \ge (1 + \delta) \mu \right) \le e^{-\delta^2 \mu / 3}$.
\end{lemma}
\begin{IEEEproof}
For any $t > 0$, we have
\begin{equation}
    \Pr\left( X \ge (1 + \delta) \mu \right) = \Pr\left( e^{tX} \ge e^{t(1 + \delta) \mu} \right)
    \le \frac{E(e^{t X})}{e^{t (1 + \delta) \mu}}.
\end{equation}
Since $E(e^{t X}) = e^{(e^t - 1)\mu}$ for a Poisson random varaible, we have
\begin{equation}
    \Pr\left( X \ge (1 + \delta) \mu \right) \le \frac{e^{(e^t - 1)\mu}}{e^{t (1 + \delta) \mu}}.
\end{equation}
Setting $t = \ln(1 + \delta) > 0$, we have
\begin{equation}
    \Pr\left( X \ge (1 + \delta) \mu \right) \le \left( \frac{e^\delta}{(1 + \delta)^{(1 + \delta)}}\right)^{\mu}.
\end{equation}
Finally, note that $\frac{e^\delta}{(1 + \delta)^{(1 + \delta)}} \le e^{-\delta^2 / 3}$ for $0 < \delta < 1$. This completes the proof.
\end{IEEEproof}



\begin{thebibliography}{00}
\bibitem{Nakamoto} Satoshi Nakamoto. Bitcoin: A peer-to-peer electronic cash system. 2008.
\bibitem{Talk} Chen Feng. Theoretical Foundation of Blockchain Technology. In \emph{16th Canadian Workshop on Information Theory}, June 2019. 
\bibitem{Tse} Vivek Bagaria, Sreeram Kannan, David Tse, Giulia Fanti, and Pramod Viswanath. Deconstructing the blockchain to approach physical limits. arXiv preprint arXiv:1810.08092 (2018). Accepted by CCS'2019.
\bibitem{UIUC} Ling Ren. 2019. Analysis of Nakamoto Consensus. Cryptology ePrint Archive, Report 2019/943. (2019). https://eprint.iacr.org/2019/943.
\bibitem{Garay} Juan Garay, Aggelos Kiayias, and Nikos Leonardos. The bitcoin backbone protocol: Analysis and applications. In \emph{Annual International Conference on the Theory and Applications of Cryptographic Techniques}, pages 281-310. Springer, 2015.
\bibitem{Pass} Rafael Pass, Lior Seeman, and abhi shelat. Analysis of the blockchain protocol in asynchronous networks. In \emph{Annual International Conference on the Theory and Applications of Cryptographic Techniques}, pages 643-673. Springer, 2017.
\bibitem{Kiffer} Lucianna Kiffer, Rajmohan Rajaraman, and abhi shelat. A better method to analyze blockchain consistency.  In \emph{Proceedings of the 2018 ACM SIGSAC Conference on Computer and Communications Security}, pages 729-744. ACM, 2018.
\bibitem{Zhao} Jun Zhao. An analysis of blockchain consistency in asynchronous networks: Deriving a neat bound. arXiv preprint arXiv:1909.06587. (2019).
\bibitem{Book} Michael Mitzenmacher and Eli Upfal. Probability and Computing: Randomized Algorithms and Probabilistic Analysis. Cambridge University Press, NY, USA, 2015. 
\end{thebibliography}
\end{document}